\documentclass[conference]{IEEEtran}
\usepackage{fancyhdr}
\usepackage{epsfig}
\usepackage{graphicx}
\usepackage{epstopdf}
\usepackage{balance}
\usepackage{caption}
\usepackage{subcaption}
\usepackage{comment}
\usepackage{color} 
\usepackage[normalem]{ulem} 
\usepackage{url}
\usepackage{algpseudocode}
\usepackage{algorithm}
\usepackage{algpseudocode}
\usepackage{pifont}
\usepackage{bbm}
\usepackage{multirow}
\usepackage{tcolorbox}
\usepackage{tikz}
\usepackage{comment}
\usepackage{placeins}
\usepackage{amsmath, amssymb}
\usetikzlibrary{arrows.meta,backgrounds,calc}

\usetikzlibrary{arrows.meta,positioning,calc,fit,backgrounds,shapes.geometric,shapes.symbols}

\definecolor{navy}{HTML}{1F4E79}
\definecolor{teal}{HTML}{2E8B57}
\definecolor{orange}{HTML}{C9791A}
\definecolor{burgundy}{HTML}{8C2D2D}
\definecolor{slate}{HTML}{4F5B66}

\begin{document}

\title{High Speed Network Processing for Scientific Computing in the Open Cloud Testbed}

\author{
    \IEEEauthorblockN{
    Miriam Leeser\IEEEauthorrefmark{1}, Michael Zink\IEEEauthorrefmark{2}, Suranga Handagala\IEEEauthorrefmark{1}, Carlos Ruben Dell'Aquila\IEEEauthorrefmark{2}
    }
    \IEEEauthorblockN{
    Verena Martinez Outschoorn\IEEEauthorrefmark{2}, Rafael Coelho Lopes de Sa\IEEEauthorrefmark{2}
    }
    \vspace{6pt}

\IEEEauthorblockA{\IEEEauthorrefmark{1}Northeastern University, Boston, MA}
\IEEEauthorblockA{\IEEEauthorrefmark{2}University of Massachusetts Amherst, Amherst, MA}
}

\maketitle

\thispagestyle{fancy} 
\lhead{} 
\rhead{} 
\chead{} 
\lfoot{\footnotesize{ SC26 Workshops, November 15-20, 2026, Chicago, Illinois, USA
\newline 979-8-3195-1221-5/26/\$31.00 \copyright 2026 IEEE}} 
\rfoot{} 
\cfoot{} 
\renewcommand{\headrulewidth}{0pt} \renewcommand{\footrulewidth}{0pt}


\begin{abstract}
This paper presents an FPGA-based SmartNIC architecture designed to address critical network challenges in data-intensive scientific computing. As scientific workflows increasingly demand high-bandwidth data movement and real-time monitoring across wide-area networks, the Open Cloud Testbed (OCT) provides an ideal platform for investigating in-network processing solutions at line rate.

Our approach focuses on implementing network visibility, optimization, and orchestration capabilities through FPGA acceleration. The proposed system leverages P4-programmable data-plane functions to support packet parsing, marking, telemetry extraction, traffic steering, and machine learning inference—all without requiring host CPU intervention or modifications to application servers.

As a concrete example, our framework integrates the SciTags standard, which enables associating network traffic with scientific context through packet marking. We implement a high-performance network bridge that counts packets by their SciTag values, as identified in the IPv6 Flow Label.

Experimental evaluation on AMD U280 FPGAs in OCT demonstrates the feasibility of this approach, showing imperceptible processing latency and line-rate packet handling, while accurately monitoring scientific data flows. 

\end{abstract}


\maketitle

\section{Introduction}

The Open Cloud Testbed (OCT)~\cite{oct-web,zink-oct21,Leeser2021octfpga,Leeser2024octfpga} has been operational since 2019, and offers AMD FPGAs, specifically U280s, VCK5000s and V80s, to researchers around the world. All of these FPGAs are directly connected to high-speed networking, enabling research on in‑network processing as well as the investigation of advanced scientific problems that involve high‑throughput data streams.

The next generation of data-intensive science will be constrained by networks as much as by compute and storage. Major computing sites are preparing for sustained wide-area traffic in the 400~Gb/s region and beyond, while scientific AI workflows are creating an additional demand for low-latency, high-bandwidth movement of training data, model checkpoints, inference inputs, and derived products. 

Improvements in computer networks are being pursued in several directions including \textit{network visibility}, \textit{network optimization}, and \textit{network orchestration}. Visibility is needed so sites and network providers can identify the owner and purpose of traffic, correlate transfers with workflows, and understand performance anywhere along the path. Optimization is needed to make more efficient use of scarce shared bandwidth through pacing, QoS, congestion-aware control, and improved handling of high-rate flows. Orchestration is needed so workflows can request elevated service when required, use reserved paths when available, and fall back gracefully to best-effort otherwise. These are not independent problems: visibility produces the labels and telemetry required for optimization, and optimization informs the orchestration policy applied to scientific workflows.

OCT provides the framework for direct experimentation on shared campus or border switching infrastructure.  We use FPGAs in OCT  that are located at the data-transfer edge, where they can inspect, classify, mark, pace, and steer traffic at line rate without requiring changes to the server and without demanding direct control of shared routers. 

To address the challenges, we propose not just a faster implementation of existing software functions, but a unifying platform for a broader set of services that are currently fragmented: SciTags packet marking and monitoring~\cite{scitags}, high-rate telemetry extraction, pacing and congestion management, QoS enforcement, dynamic IP/VLAN handling for orchestration, and compact AI/ML inference close to the packets. 

Prior work has shown that P4-defined data-plane functions can be compiled to FPGA hardware at line rate, that FPGA coprocessors can support flexible telemetry at data-plane rates, and that real-time in-network ML inference on P4-programmable FPGA SmartNICs is feasible for tasks such as QoS prediction and anomaly detection~\cite{han2023framework,Bal2024cnert,han2024memory,han2024extracting}. FPGA acceleration provides low latency, deterministic packet access, and a deployable hardware target for compact models that can act faster than host-based control loops.

The goals of this project are to develop a flexible firmware-centric implementation of high-bandwidth network functions on AMD/Xilinx SmartNICs, leveraging P4-based packet-processing methods with support for packet parsing, marking, telemetry, local steering, and control hooks for ML inference.

The research presented in this paper addresses network visibility by demonstrating real-time insertion, inspection, and monitoring of SciTags-compatible IPv6 packet markings without interfering with application-level servers, thereby creating science-aware telemetry suitable for multimodal and AI datasets.  Specifically,  this research monitors SciTags-compatible IPv6 packet markings on FPGAs directly connected to the network.  Our experiments demonstrate that the FPGA processes the traffic and produces identical results as bypassing the FPGA and doing all processing on the processor.  Adding an FPGA in the flow of data introduces minimal extra latency and allows for processing at network line rates.   These preliminary results demonstrate the potential for in-network data processsing by FPGAs in OCT.

\section{Background}
\label{sec:background}

\subsection{SciTags \& SENSE}

The computing models of next-generation physics experiments will couple distributed storage systems, high-performance computing facilities, and analysis platforms through heterogeneous wide-area networks. Aggregate interface counters and sampled flow records will therefore no longer provide sufficient information. Monitoring will need to be end-to-end and aware of the scientific context of each transfer. Scientific Network Tags (SciTags) address this semantic gap by associating network traffic with standardized community and activity identifiers, conveyed through packet marking~\cite{Attebury2025SciTags,scitags}.

These identifiers allow measurements of throughput, packet loss, latency, route changes, and congestion to be correlated with a particular experiment, production campaign, or analysis activity. Programmable network-interface cards built around FPGAs such as SmartNICs can implement tag insertion and parsing, precise timestamping, per-class counters, and compact streaming summaries directly in the data path. Such an implementation would provide deterministic telemetry with limited involvement of the host CPU, while exporting summarized measurements to software services for correlation, anomaly detection, and longer timescale diagnosis.

Network orchestration can ultimately elevate connectivity and bandwidth to schedulable resources alongside computing and storage. A scientific workflow should be able to express an intent (endpoints, requested bandwidth, priority, lifetime, and isolation requirements) and have the infrastructure provision and monitor the corresponding service across multiple possibly distributed administrative domains. The Software-Defined Network for End-to-end Networked Science at the Exascale (SENSE) research project provides a model-driven, intent-based architecture for delivering such end-to-end services~\cite{Monga2020SENSE}. Subsequent work has applied this approach directly to Large Hadron Collider (LHC) data distribution~\cite{Balcas2024AutomatedNetworkServices}. More recently, the Data Movement Manager prototype has connected the LHC dataset manager Rucio with SENSE, translating transfer priorities into bandwidth allocation~\cite{Arora2025DMM}. FPGA-based endpoints and domain-edge devices can serve as the orchestrator enforcement layer, matching SciTags and reservations to queues and returning high-fidelity telemetry.  This research monitors SciTags-compatible IPv6 packet markings on FPGAs directly connected to the network.  

\subsection{P4 on FPGAs}

One advantage of using an FPGA as a SmartNIC is its programmability at the hardware level. The FPGA allows the developer to customize the packet paths and create high performance computing units. There are two open source FPGA-based NIC designs that provide  the fundamental shell for FPGA-based SmartNICs: the OpenNIC Project~\cite{opennic} and Corundum~\cite{forencich2020fccm}.  They both provide the hardware designs for the NIC and corresponding Linux drivers. Both of them consist of the Media Access Control (MAC) and Physical Interface Layers on one side to communicate with the Ethernet connection, and Direct Memory Access to connect with the host CPU.  OpenNIC is optimized for AMD FPGAs which is why we employ it for this project.  In addition to providing tools for network programming, Siddhartha, et al.\ present a project based on OpenNIC~\cite{p4demo_dnn} that provides a SmartNIC framework tailored for deep neural networks. 

\subsection{OCT}

\label{sec:Open Cloud Testbed}

The Open Cloud Testbed (OCT)~\cite{zink-oct21, Leeser2021octfpga,Handagala2021cnert,han2023framework,Bal2024cnert,Leeser2024octfpga} is an NSF-funded CIRC project whose goal is to provide bare-metal servers to the research community. In contrast to other testbeds that offer CPUs and GPUs, OCT offers a large set of FPGAs directly connected to the network that can be programmed by testbed users~\cite{Leeser2021octfpga, Leeser2024octfpga}.  

During the initial phase, OCT consisted of eight AMD Alveo U280 accelerator cards configured with direct network connectivity. Pre-compiled TCP and UDP network stack binaries for the FPGAs were provided to support experiments involving network-attached FPGAs. User support primarily focused on the AMD Vitis flow, which enabled users to build their projects on top of a pre-programmed FPGA shell provided by AMD. Users could gain access to these FPGAs by joining a designated CloudLab project, with FPGA resources provisioned through a CloudLab profile. Support for projects involving custom shells such as OpenNIC and Corundum remained restricted during this stage, and access for those users was granted through a separate project. 

Subsequent expansions significantly increased the capabilities of OCT through the addition of sixteen more U280 accelerators, bringing the total to 24. Along with this expansion, a unified workflow was created to support both Vitis and custom flows, eliminating the need to maintain a separate project for users who use the custom flow. Supporting Vivado users through a separate project required manual administrative intervention to execute factory resets between experiments, which created significant management challenges. To resolve this, an automated workflow was implemented to factory reset the FPGAs independent of the workflow. Specifically an FPGA is programmed with a golden shell at experiment termination. As a result, the next user always receives a clean bare-metal FPGA free of any prior shell configuration. This approach successfully supports both Vivado and Vitis flows within a single project framework with minimal manual intervention.

\begin{figure}
    \centering
    \includegraphics[width=1\linewidth]{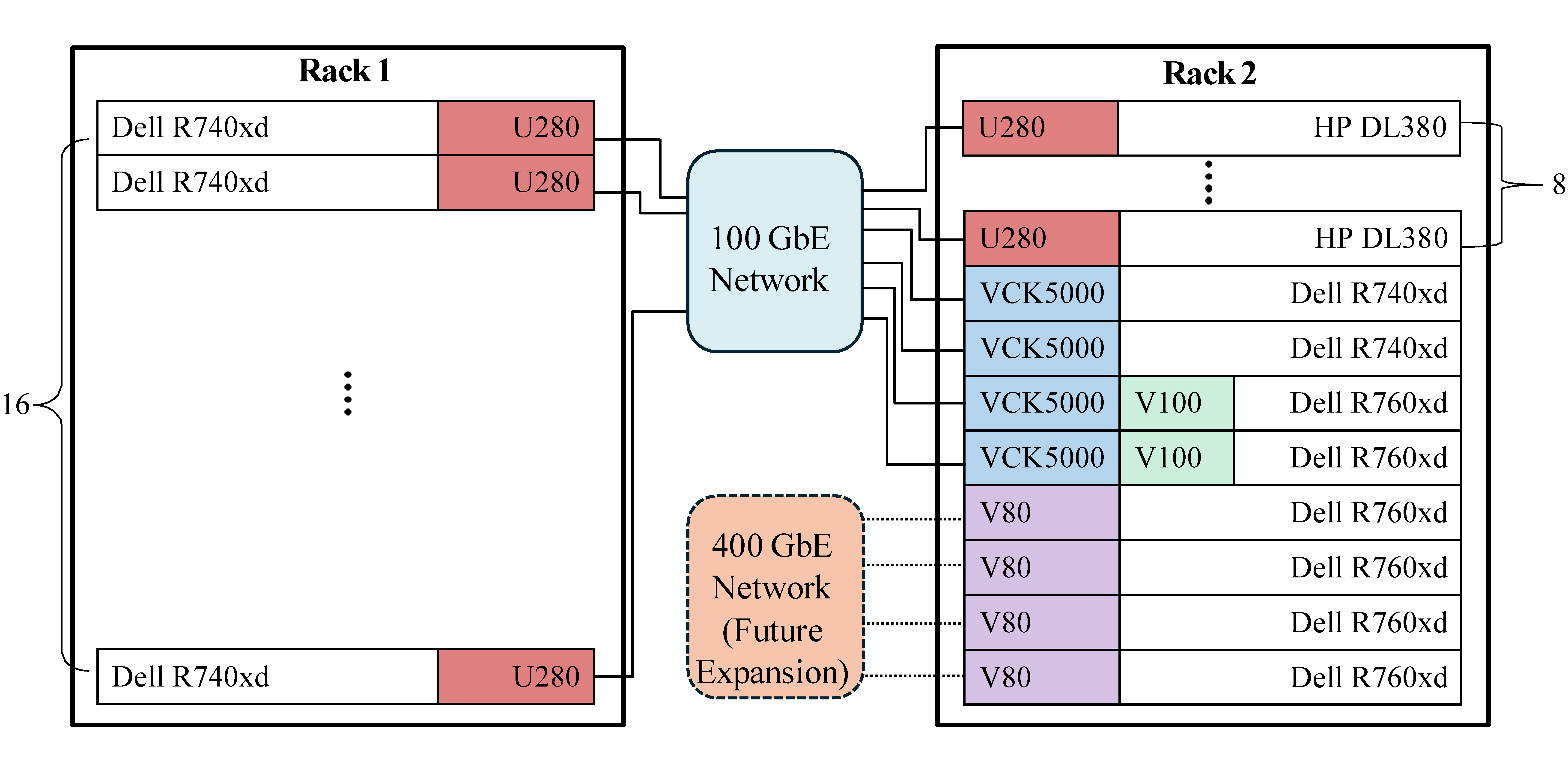}
    \caption{Network connections for FPGAs in OCT}
    \label{fig:network-connections}
\end{figure}

While the U280 has long served as the backbone of the FPGAs in OCT, our infrastructure has steadily evolved to support a diverse range of workloads including machine learning and AI. To address these emerging applications, OCT has incorporated AMD VCK5000 and V70 accelerators that are based on the Versal Adaptive SoC architecture. These devices come with specialized AI engines that are capable of accelerating modern workloads such as real-time object detection, image classification, and video analytics.

\begin{figure*}[!htb]
    \centering
\includegraphics[width=\linewidth]{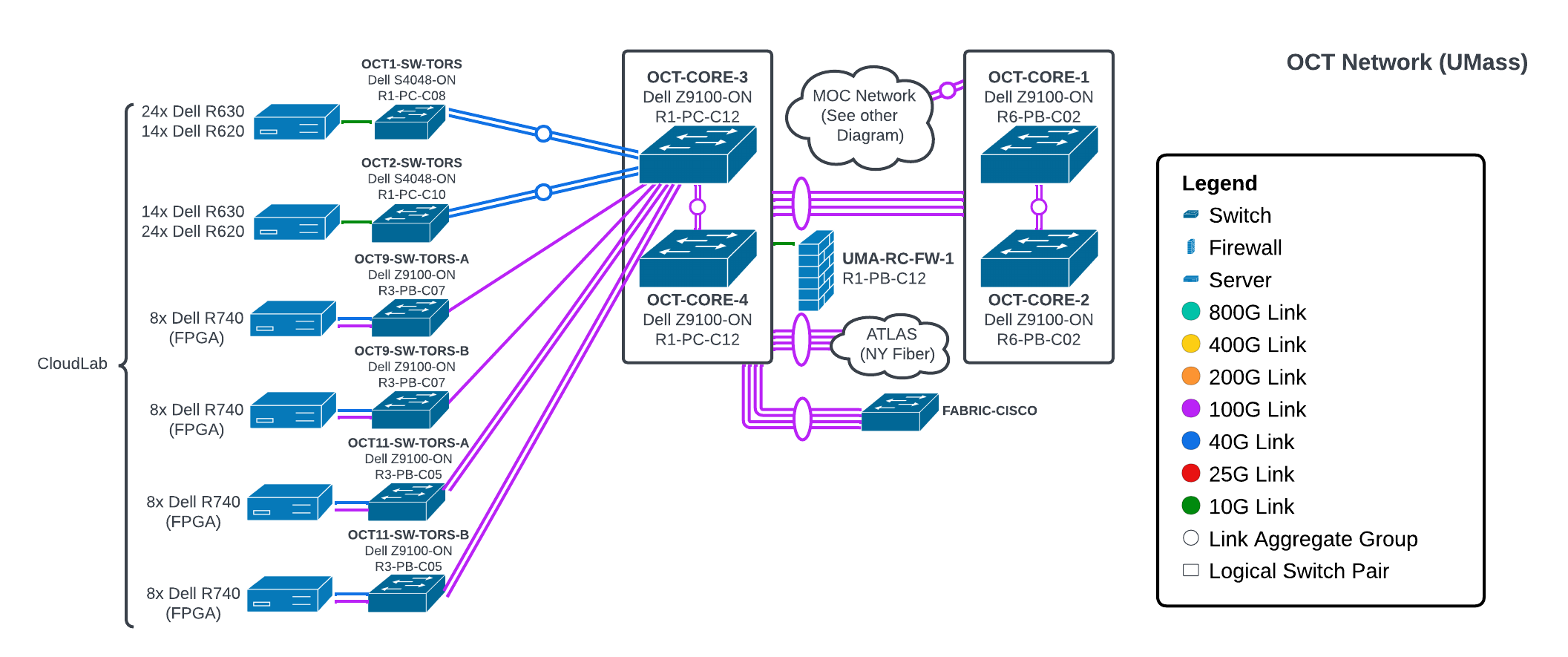}
    \caption{Network topology of the Open Cloud Testbed in the MGHPCC. }
    \label{fig:MGPCC_topo}
\end{figure*}

\begin{table*}[!ht]
\centering
\caption{Comparison of AMD Accelerator Cards in OCT: Alveo V80, VCK5000, and U280}
\label{tab:accelerator-comp}
\begin{tabular}{p{3cm} p{3.8cm} p{3.8cm} p{3.8cm}}

\textbf{Feature / Specification} & \textbf{AMD Alveo V80 \newline (Versal HBM Series)} & \textbf{AMD VCK5000 \newline (Versal AI Core Series)} & \textbf{Xilinx Alveo U280 \newline (UltraScale+ FPGA)} \\

\textbf{Architecture} & 7nm Versal Adaptive SoC & 7nm Versal ACAP & 16nm UltraScale+ FPGA \\

\textbf{Primary Focus} & High-bandwidth memory, massive I/O, \& networking & AI Inference, DSP, and vector processing & General-purpose acceleration, database analytics, ML inference  \\

\textbf{Compute Engines} & Adaptive Logic/DSP/ ARM Subsystem & Vector AI Engines + Adaptive Logic/DSP & UltraScale+ Logic Cells \& DSP Slices \\

\textbf{Memory Architecture} & 32 GB HBM2e ($\approx$820 GB/s) + 4 GB DDR4 & 16 GB Off-chip DDR4 ($\approx$102.4 GB/s) & 8 GB HBM2 ($\approx$460 GB/s) + 32 GB DDR4 \\

\textbf{Embedded Processing} &  Dual-core Arm Cortex-A72 APU, Dual-core Arm Cortex-R5F RTPU & Dual-core Arm Cortex-A72 APU, Dual-core Arm Cortex-R5F RTPU & None \\

\textbf{Networking Interfaces} & 4x200G QSFP56 (800 Gb/s total) & 2x100G QSFP28 (200 Gb/s total) & 2x100G QSFP28 (200 Gb/s total) \\

\textbf{PTP} & Native hard-block IEEE 1588 TSU & No native support & No native support \\

\textbf{Inline Security} & Hardened 400G cryptographic engines (IPsec \& MACsec) & None & None \\

\textbf{Host Interface} & PCIe Gen5x8 or Gen4x16 & PCIe Gen4x8 or Gen3x16 & PCIe Gen4x8 or Gen3x16 \\

\end{tabular}
\end{table*}

Currently, OCT provides 34 AMD FPGAs: 24 Alveo U280, 4 Alveo VCK5000, 4 Alveo V70, and 2 Alveo V80. OCT is connected to the US ATLAS Northeast Tier 2 network and to ESNet at 400 Gb/s, as well as to the NSF FABRIC national research infrastructure. Fig.~\ref{fig:network-connections} shows the networking connections within OCT, and Fig.~\ref{fig:MGPCC_topo} expands on the details of the MGHPCC topology.  Table~\ref{tab:accelerator-comp} compares the capabilities of the various different accelerators available.  The V80s are described in more detail in the next section.

\subsection{V80s in OCT}
\label{sec:v80}

AMD V80 is a high-performance reconfigurable compute accelerator card built on the 7nm Versal Adaptive SoC architecture, designed for heterogeneous data center environments. It provides 32 GB of HBM2e memory and 820 GB/s peak bandwidth to that memory. This significantly increases memory throughput for bandwidth-intensive workloads while reducing memory bottlenecks. Additionally, V80s come with 4x200G QSFP56 optical network ports that provide 800 Gb/s of total aggregate network bandwidth for applications such as distributed computing clusters and high-speed data transfer. These high-capacity interfaces allow the accelerator to transmit massive data streams directly at line rate. 

Compared to the previous generation UltraScale+ architecture used in the U280s, Versal Adaptive SoC architecture integrates several functions as dedicated hardened IP blocks. Features that previously required implementation using FPGA programmable logic are now provided as hard IPs, reducing LUT and routing resource utilization, and improving performance and power efficiency. Examples of these hardened IPs include cryptographic accelerators and IEEE 1588 PTP engines. Rather than implementing these functions in the FPGA fabric, designers can instantiate the corresponding hard IPs through AMD's development tools, which automatically integrate the hardened resources with the surrounding FPGA logic. As a result, more FPGA fabric remains available for implementing user logic.

\subsection{Build Machines}
\label{sec:bm}

During the early stage of OCT, Build tools were hosted within the Mass Open Cloud (MOC), and subsequently on the New England Research Cloud (NERC), which allowed users to compile their bitstreams and manage software dependencies independent of the CloudLab host nodes. While keeping build and deployment tasks in separate environments provided a clean boundary, it also introduced several disadvantages, such as the administrative overhead of maintaining two separate user accounts, and the need to synchronize compilation artifacts such as FPGA bitstreams and host executables between two different network domains. To streamline the development workflow, the build tools were subsequently migrated to CloudLab, allowing the build and host nodes to reside within a single environment.  

OCT currently includes dedicated bare-metal build machines packed with over $2\text{TB+}$ of aggregate memory, 32 cores of AMD Ryzen 5950X, and 128 cores of AMD EPYC 7C13 CPUs to support users build/emulate their FPGA projects. By using a CloudLab profile, users can leverage these resources and provision virtual machines by specifying the number of CPU cores and memory they need for their workloads.  Tutorials and environments for the V80s and for OCT in general are available at \cite{oct-web}.

\section{Applications}
\label{sec:apps}

\begin{figure*}
    \centering
    \includegraphics[width=0.8\linewidth]{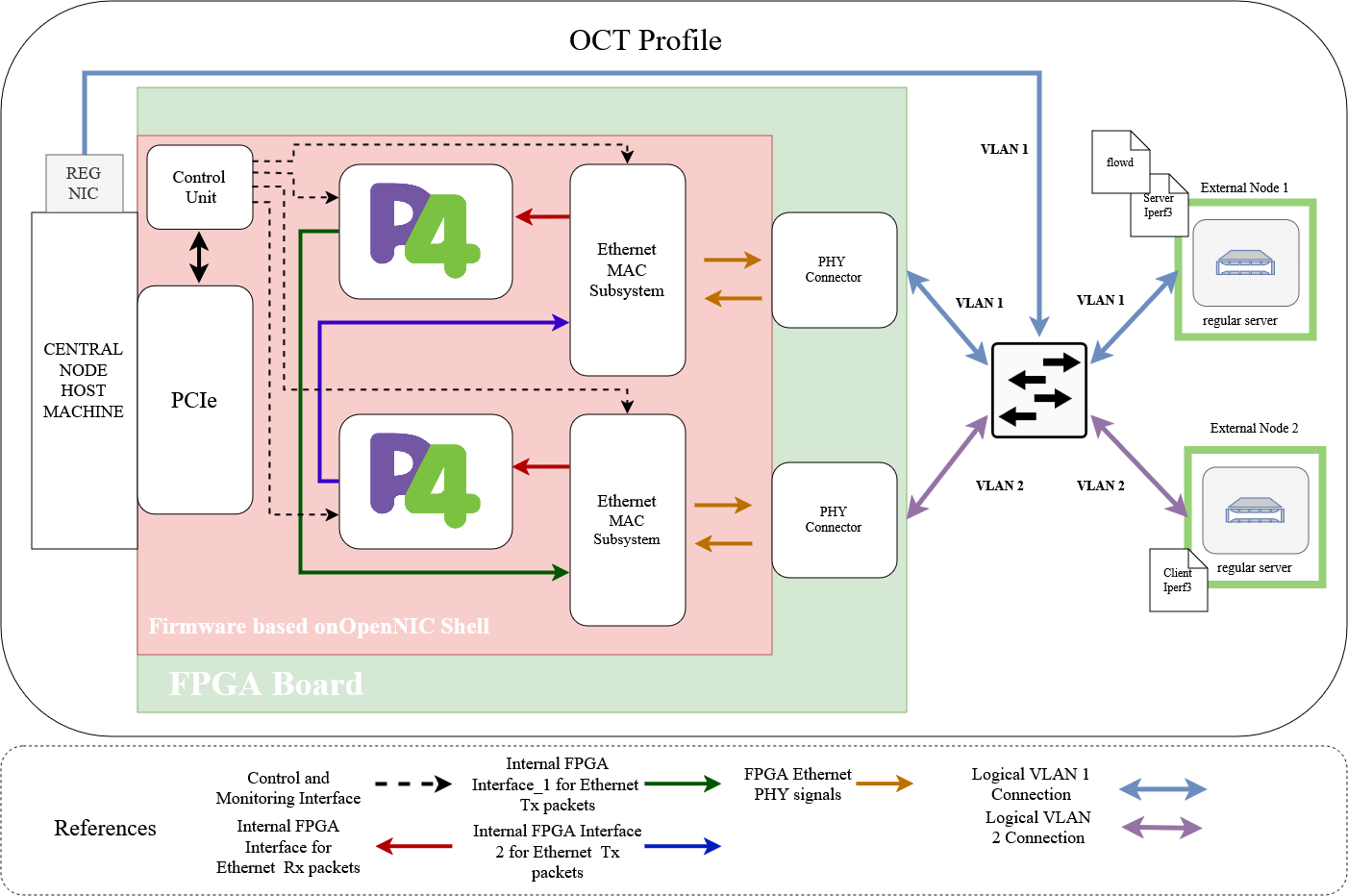}
    \caption{Network telemetry based on Scitags, and tested on the OCT platform}
    \label{fig:network-telemetry}
\end{figure*}

We have implemented and evaluated a preliminary network telemetry architecture. It is based on FPGA hardware that counts the IPv6 packets exchanged between two networks according to their Scitag value. Figure~\ref{fig:network-telemetry} shows a general diagram of the solution; more details of the components are presented below. Implementation and evaluation were performed at the Massachusetts High Performance Compute Center (MGHPCC), which houses the ATLAS NET2 cluster and OCT.

To allocate the resources required for our evaluation, an OCT profile is associated with a Python script that programmatically defines the server nodes, their network interconnections through switches, and VLAN settings. The external nodes exchange iperf application packets marked with Scitags. Their regular NICs are associated with different VLAN IDs for emulating two systems, such as scientific data storage and data management systems working on different networks ~\cite{scitags}. The central node is responsible for bridging the traffic between the external ones. It targets one of the FPGA boards supported. It implements the bridge and counts the packets according to their Scitag value. Additionally, the central node has a regular NIC connected to one of the VLANs. Its purpose is to conduct a throughput test against the corresponding external node, without any additional device in the datapath beyond the network switch. The results are used as a reference to compare the throughput obtained from the test between the two external nodes, when packets pass through the FPGA or not. This allows quantification of performance reduction introduced by the proposed telemetry system.

\textit{Flowd} is a software package developed by the Scitags initiative group~\cite{scitags}. It comprises a set of plugins and back-end services for generating and logging Scitags on traffic generated by scientific applications such as network performance tests like iperf, or storage applications. In the context of this work, the plugin for IPv6 packets is used. It runs as a Linux kernel module, marking packets with Scitag values.

The FPGA design has two relevant components for the proposed solution. The first is the OpenNIC Shell framework which provides the PCIe interface between the host system and the FPGA board, Ethernet MAC blocks, bus interfaces infrastructure, and user logic. The second supports the integration of custom network functions, which is implemented based on the AMD Vitis Networking P4 (VNP4) IP Core in this work. The VNP4 compiles a P4 program in RTL design, and two cores compiled with the same P4 program are instantiated. Each of these receives Ethernet packets from one of the MAC blocks for processing according to the P4 program, and then forwards the packet to the other port to complete the bridge operation.

\begin{figure}
    \centering
    \includegraphics[width=1\linewidth]{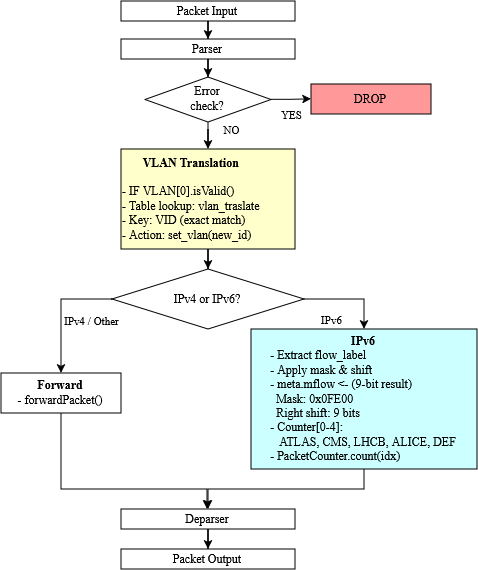}
    \caption{P4 program diagram for network telemetry application}
    \label{fig:p4-program-network-telemetry}
\end{figure}

The P4 program is summarized in Figure~\ref{fig:p4-program-network-telemetry}. There are two tables, and the key for each should exactly match one of the entries loaded at runtime.  As the key is in both cases $<=$ 16 bits, the tables are implemented with Block RAMS (BRAM) on the FPGA.  The first table is used for VLAN translation, so the key is the VLAN ID values of the packet header which needs to be swapped, which is important for the packet bridging between the two networks. The second table is associated with packet counters. Its key is the experiment information (9-bits) of the Scitag value contained in the Flow Label (20-bits) field of the IPv6 packet header. Standard P4 logical operations such as bit masking and shift are used for extracting the experimental information. As soon as it matches with one of the table entries loaded at runtime, the corresponding experiment’s counter is incremented.

Host system software was developed to load the table entries for the VLAN translation and packet counters. The counters are  associated with experimental information from Scitags. The software also periodically reads the counters' values and saves this information in a local database. 

OCT supports two models of NICs for the external node configuration. One provides 40 Gb/s of nominal throughput and an MTU (Maximum Transmission Unit) of 1500 bytes, while the second provides 100 Gb/s and an MTU of 9000 bytes, respectively. One of these options is chosen during experiment creation, and both of them were used to evaluate the performance of the proposed architecture in terms of throughput degradation.

The performance test procedure comprises two measurements. The first one is the reference throughput between two regular NICs of the experiment nodes, 40 Gb/s in the first case and 100 Gb/s in the second one. This is then compared with the second measurement, which is obtained from the resulting throughput between the same models of NICs used for the reference measurement, but with all packets passing through the FPGA of an Alveo U280 card. This card acts as a bridge between the two NICs and runs the firmware implementing the proposed telemetry system. The throughput measurement is performed by running the iperf3 application. The results are summarized in Table~\ref{tab:throughput}.

\begin{table}[ht]
\centering
\caption{Network throughput achieved by the reference implementation and the FPGA-based bridge.}
\label{tab:throughput}
\begin{tabular}{|c|l|c|c|c|}
\hline
\textbf{NIC} & &
\shortstack{\textbf{Sender}\\(Gb/sec)}
&
\shortstack{\textbf{Receiver}\\(Gb/sec)}
&
\shortstack{\textbf{Retransmission}\\(packets)}
\\
\hline

\multirow{2}{*}{40Gb/s}
& Reference
& 37.5
& 37.0
& 2021
\\
& FPGA-based bridge
& 37.5
& 37.0
& 2039
\\
\hline

\multirow{2}{*}{100Gb/s}
& Reference
& 95.6
& 95.1
& 14958
\\
& FPGA-based bridge
& 96.1
& 95.3
& 13719
\\
\hline

\end{tabular}
\end{table}

The second test  verifies that the FPGA counters are working properly. The resulting packets are saved and post-processed offline. The number of packets from the offline analysis is compared with those from the FPGA and host system application. The result are presented in Table~\ref{tab:event_counters}.  

\begin{table}[ht]
\centering
\caption{Comparison of event counters obtained from the FPGA and the PCAP file.}
\label{tab:event_counters}
\begin{tabular}{|l|r|r|}
\hline
\textbf{Experiment Counter} & \textbf{FPGA} & \textbf{post-processed} \\
\hline
ALTAS  & 854311 & 854311 \\
CMS    & 744050 & 744050 \\
LHCb   & 765889 & 765889 \\
ALICE  & 841103 & 841103 \\
\hline
\end{tabular}
\end{table}

Table~\ref{tab:event_counters} show that packets processed through the FPGA provide exactly the same result as those that bypass the FPGA. Our throughput results  at both  40Gb/s and at 100Gb/s show that the FPGA processing of network packets introduce an imperceptible amount of delay.  In addition, the FPGA based design required fewer retransmissions at high data rates.   

In the future we plan to move the entire processing to the V80s and connect to a 400Gb/s network interface.  At those speeds,  the FPGA should be able to handle the network traffic at line rate and to preprocess the data before handing it to the processor.

\section{Conclusions}
\label{sec:conclusions}

This work establishes the foundation for an FPGA-based SmartNIC architecture in which network visibility, optimization, and orchestration are implemented as coordinated data-plane and control-plane capabilities. On the data-plane side, we have demonstrated high-throughput firmware support for the insertion and monitoring of SciTags, together with programmable VLAN manipulation. These functions allow scientific traffic to be identified, measured, and associated with network services directly in the packet-processing path, without requiring host-CPU intervention for every packet. Support for Differentiated Services Code Point (DSCP) marking and processing is currently under development and will extend this functionality to traffic prioritization and quality-of-service policies. Collectively, these developments provide a substantial implementation of the required P4-programmable data plane. Packets can be classified and modified in FPGA firmware at the bandwidths required by data-intensive physics applications.

The next stage of the project is the development of the corresponding control plane. Rather than applying SciTags, VLAN, and DSCP operations as static global configurations, the control plane will manage these functions according to configurable policies, initially at the granularity of IPv6 prefixes or subnets. It will maintain the mapping between scientific activities and their required data-plane actions, configure SciTags identifiers and counters, associate SENSE-provisioned services with the appropriate VLANs, and apply DSCP policies to selected traffic classes. A programmatic interface will expose these capabilities to external monitoring systems and network orchestrators, while also providing access to firmware-generated telemetry, including per-activity packet and byte counters. This separation preserves the deterministic performance of FPGA packet processing while allowing network policies to be created and retired dynamically in software.

This project also demonstrates the ability of OCT to flexibly support scientific applications and processing in the network.  We welcome users from other application domains who can make use of this open, flexible and configurable platform.  The experiments described here have been conducted on AMD U280 FPGAs.  We plan to extend them to the AMD V80s which offer the capability of much higher speed networking.

\section*{Acknowledgment}
This work makes use of the Open Cloud Testbed, funded by the National Science Foundation (NSF) under grants 1925464, 1925658, 2130907, 2130891, 2319962, 2332423 and support from the Department of Energy (DOE) award DE-SC0010004. We would like to thank AMD for their generous donations.  


\bibliographystyle{acm}
\bibliography{bib}

\end{document}